\documentclass[aps,prb,twocolumn,showpacs,preprintnumbers,amsmath,amssymb,superscriptaddress]{revtex4}
\usepackage{color}
\usepackage{graphicx}
\usepackage{dcolumn}
\usepackage{bm}
\usepackage[hypertex]{hyperref}
\newcommand{\nc}{\newcommand}
\nc{\be}{\begin{eqnarray}}
\nc{\ee}{\end{eqnarray}}
\nc{\bea}{\begin{eqnarray}}
\nc{\eea}{\end{eqnarray}}
\nc{\bean}{\begin{eqnarray*}}
\nc{\eean}{\end{eqnarray*}}
\nc{\mb}{\mbox}
\nc{\rnc}{\renewcommand}
\nc{\vk}{\mb{\bf k}}
\nc{\vp}{\mb{\boldmath$p$}}
\nc{\rr}{\mb{\boldmath$r$}}
\nc{\RR}{\mb{\boldmath$R$}}
\nc{\vz}{\hat {\mb{\bf z}}}
\nc{\vj}{\mb{\boldmath$j$}}
\nc{\vg}{\mb{\boldmath$g$}}
\nc{\x}{\mb{\boldmath$x$}}
\nc{\A}{\mb{\boldmath$A$}}
\nc{\va}{\mb{\boldmath$a$}}
\nc{\vq}{\mb{\boldmath$q$}}
\nc{\vn}{\mb{\boldmath$n$}}
\nc{\vs}{\mb{\boldmath$\sigma$}}
\nc{\vt}{\mb{\boldmath$\tau$}}
\nc{\vpi}{\mb{\boldmath$\pi$}}
\nc{\nab}{\bm{\nabla}}
\nc{\X}{\sf x}

\begin{document}


\title{
Spin-Orbit Interaction Enhanced Fractional Quantum Hall States\\
in the Second Landau Level
}

\author{Toru Ito}
\affiliation{
Department of Physics, Tohoku University, Sendai, 980-8578, Japan
            }
\author{Kentaro Nomura}
\affiliation{
Department of Physics, Tohoku University, Sendai, 980-8578, Japan
            }
\affiliation{
Correlated Electron Research Group (CERG), RIKEN-ASI, Wako 351-0198, Japan
            }
\author{Naokazu Shibata}
\affiliation{
Department of Physics, Tohoku University, Sendai, 980-8578, Japan
            }

\date{\today}

\begin{abstract}
We study the fractional quantum Hall effect at $\nu = 7/3 $ and 5/2 in the presence of the spin-orbit interaction,
 using the exact diagonalization method and the density matrix renormalization group (DMRG) method in a spherical geometry.
 Trial wave functions at these fillings are the Laughlin state and the Moore-Reed-Pfaffian state.
 The ground state excitation energy gaps and pair-correlation functions at fractional filling factor $\nu =7/3$ and 5/2
 in the second Landau level are calculated.
 We find that the spin-orbit interaction stabilizes the fractional quantum Hall states.
\end{abstract}

\def\runauthor{T. I\sc{to} \textit{et al}.}

\maketitle

The fractional quantum Hall (FQH) effect is a remarkable many-body phenomenon observed
in two dimensional electron gases (2DEG) in a strong perpendicular magnetic field\cite{Tsui1982}.
The Laughlin state and it's hierarchy theory describe the fractional quantum Hall (FQH) effect at odd-denominator
filling factors successfully.\cite{Laughlin1983,Haldane1983} The feature of Laughlin state is the appearance of quasiparticle
excitations with fractional charge and fractional statistics called anyon.\cite{Halperin1983}
On the other hand, the Laughlin state does not include an even-denominator state.
Therefore the quantum Hall plateau observed at filling factor $\nu =5/2$ (second Landau level even dominator) poses a
special challenge.

There are some candidates for the ground state (GS) at $\nu = 5/2$.
Particularly, the Moore-Read Pfaffian (Pf) state is believed to be a strong candidate for fractional quantum
Hall state at this filling.\cite{Moore1991,Greiter1992}
The Pfaffian wave function is written by \\
\begin{equation}
\Psi_{\rm Pf}={\rm Pf}\Big(\frac{1}{z_{i}-z_{j}}\Big)\prod _{i<j}(z_{i}-z_{j})^{2} e^{-\frac{1}{4\ell_B^2}\sum_i|z_i|^2},
\end{equation}
here $z_{i} = x_{i} -iy_{i}$, the Pfaffian factor is defined by\\
$
{\rm Pf}M_{i,j}=\frac{1}{2^{N/2}(N/2)!}\sum _{\sigma\in S_{N}}{\rm sgn}\sigma\prod ^{N/2}_{k=1}M_{\sigma(2k-1)\sigma(2k)} ,
$
for an $N \times N$ antisymmetric matrix whose elements are M$_{ij}$; S$_{N}$ is the group of permutations of $N$ objects.
The Pfaffian state is totally antisymmetric for even denominator, so could describe electrons.\cite{Note_APf}

The occurence of the Pf state is very exciting since the quasiparticle excitations in this state
has non-abelian statistics (Laughlin state excitations are abelian anyons).
The non-abelian topological phase suggests potential applications toward quantum computing.
The quantum gates constructed from abelian anyons are very limited.\cite{Averin2002}
Non-Abelian anyons, on the contrary, are much more useful to the
topological quantum computer,\cite{Kitaev2003} since the braiding of non-Abelian anyons induces noncommuting processes,
one can construct various quantum gates from non-Abelian representations of braid group.
Therefore, determining the nature of $\nu =5/2$ state is of greater urgency, beyond
the conventional FQH physics.

 Unfortunately,  experiments showed that the energy gap
at the $\nu = 5/2$ is very small and then the FQH state at this filling is fragile\cite{Willet1987,Eisenstein2002}. Numerical studies indicate the fact that the Pf state exist between the
compressible stripe state and the composit fermion liquid (CFL) state, in the phase diagram.\cite{Morf1998,Wang2009,Storni2010}
These studies showed the $\nu =5/2$ ground state is near of the stripe phase and, extra short range
interaction changes Pf state into stripe state.

In this paper we show that the Pf state could be more stable by moving the point away from the transition point, and
the spin-orbit interaction has good influence on stability of the pfaffian state. A large number of theoretical studies have been devoted to investigate the FQH states in 2DEG, 
however, the effect of the spin-orbit interaction on the FQH state has not been studied enough yet.
In this paper we investigate the FQH state in the presence of the spin-orbit interaction by using numerical method.
In 2DEGs, the spin-orbit interaction has two relevant contributions, the Rashba term and the Dresselhaus term.\cite{Rashba1960,Dresselhaus1992}

The Rashba coupling Hamiltonian is described by
\begin{equation}
H_{R}=\lambda(\Pi_{x}\sigma^{y}-\Pi_{y}\sigma^{x}) .
\end{equation}
The Dresselhaus Hamiltonians is described by
\begin{equation}
H_{D}=\lambda'(\Pi_{x}\sigma^{x}-\Pi_{y}\sigma^{y}) .
\end{equation}
Here, the spin-orbit coupling parameter  $\lambda$ is linearly dependent on the expectation value of the
interface electric field  $\langle E_{z}\rangle$  at 2DEG, this relation is described by\cite{Nitta1997}
\begin{equation}
\lambda \sim  \langle E_{z}\rangle,
\end{equation}
here $\langle E_{z} \rangle$ is controlled by the gate voltage.
Rashbha spin-orbit coupling arises from the structure inversion asymmetry of the quantum well,
and the Dresselhause term stems from the bulk-inversion asymmetry of semiconductor material.
The coefficient $\lambda$' of
the Dresselhause term is fully determined by the geometry of the heterostructure, while the Rashba coefficient $\lambda$ can be 
varied by an electric field across the well.
Lommer et al. have theoretically pointed out that the Rashba mechanism becomes dominant in
a narrow gap semiconductor system.\cite{Lommer1988} Later, Luo et al. have experimentally shown that the Rhashba
term is dominant for spin splitting in an InAs based heterostructure.\cite{Luo1990} We take account of only
the Rashba term in this paper.\\
Eigenstates of the single-particle Hamiltonian including the Rashba term
can be written in the form
\begin{equation}
 | \psi \rangle = \cos\Big(\frac{\pi}{2} \alpha\Big)| n=1 \rangle + \sin\Big(\frac{\pi}{2}\alpha\Big)| n=0 \rangle,
\end{equation} 
where $| n=1\rangle$ is a second Landau level (SLL) eigenfunction, $|n=0\rangle$ is a lowest Landau Level (LLL)
eigenfunction, cos($\alpha$) is written by,\cite{Schliemann2003}
\begin{equation}
\\
  \cos\Big(\frac{\pi}{2}\alpha\Big) = \Big(\frac{1}{2}+{\frac{\hbar\omega_{c}/4}{\sqrt[]{2n\lambda^{2}m\hbar\omega_{c}+(\hbar\omega_{c})^{2}/4}}}\Big)^{1/2} .
\\ 
\end{equation}
From eqs. (4) and (6),
we find that the mixing parameter $\alpha$ increases from $\alpha = 0$ with the increase of the perpendicular electron field.
In the limit of strong spin-orbit coupling or weak magnetic field, the value of $\alpha$ is 0.5
($\cos(\frac{\pi}{2}\alpha)$= $1/\sqrt{2}$).
This limit corresponds to the $n=1$ LL of graphene.\cite{Shibata2008} 
We note that since the GS at $\nu=1/2$ in the LLL($\alpha$ = 1.0) is shown to be the CFL state,\cite{Rezayi1994}
the Pf state will be stabilized by the weak spin-orbit interaction.

To study many-body problems, we study the projected many body interaction
Hamiltonian onto a certain Landau Level which is written as,\cite{Haldane1983}
\begin{equation}
 H=\frac{1}{L^{2}}\sum _{i<j}\sum _{\textrm{\boldmath $q$}}
V(q)e^{-\frac{q^2}{2}}[F_{n}(q)]^{2}
e^{i\textrm{\boldmath $q$}\cdot(\textrm{\boldmath $R$}_{i}-\textrm{\boldmath $R$}_{j})},
\end{equation}
where $R_{i}$ is the guiding center coordinate of the $i$th electron, which satisfies the commutation relation,
 $[R^{x}_{j},R^{y}_{k}]=i\ell_B^{2}\delta_{jk}$, and $V(q)=2\pi\frac{e^{2}}{\epsilon q}$ is the Fourier transform of 
the Coulomb interaction. In this work, the magnetic length $\ell_B$ is set to be 1, and we take ${e^{2}}/{\epsilon\ell_B}$ as
units of energy scale. We omit the component at $q=0$, which is canceled by uniform positive background charge.
We assume that the width of the wave function perpendicular to the two dimensional plane is sufficiently small
compared with the magnetic length.\cite{Michael2008}
 The relativistic form factor in the $n$th LL is
written as
\begin{equation}
F_{0}(q) = L_{0}(\frac{q^{2}}{2}) ,
\end{equation}
and
\begin{equation}
F_{n\geq 1}(q) = \cos^{2}(\frac{\pi}{2}\alpha)L_{n}(\frac{q^{2}}{2})+ \sin ^{2}(\frac{\pi}{2}\alpha)L_{n-1}(\frac{q^{2}}{2}) ,
\end{equation}
here, $L_{n}(x)$ are the Laguerre polynomials.

In this paper, we investigate the effects of the spin-orbit interaction on the Laughlin state and the Pfaffian state
 by applying this projected Hamiltonian.
We calculate the exact wave functions of the ground state for several values of $\alpha$, using the exact
diagonalization for $N \leq $14,\cite{Haldane1985,Fano1986,Morf2002} and the density matrix renormalization group (DMRG) method for $N \leq $18\cite{Shibata2001,White1992}, in
the spherical geometry. The DMRG method is a real space renormalization group method combined with a exact
diagonalization method. The DMRG method provides low-energy eigenvalues and corresponding
eigenvectors of Hamiltonians within a restricted number basis states.
The accuracy of the results is systematically controlled by the truncation error, which is smaller than
0.0003 in the present calculation.

In the spherical geometry, it is convenient to write the Hamiltonian as\cite{Haldane1983}
\begin{equation}
H^{n}= \sum_{i<j}\sum_{m}V^{n}_{m}P_{ij}[m] ,
\end{equation}
where $P_{ij}[m]$ projects onto states in which particles $i$ and $j$ have the relative angular momentum $\hbar$$m$,
and $V_{m}(n)$ is their interaction energy in the $n$th LL. Using the above form factors, the pseudopotentials are given by
\begin{equation}
V^{n}_{m}=\int ^{\infty }_{0}\frac{dq}{2\pi}qV(q)e^{-q^{2}}[F_{n}(q)]^{2}L_{m}(q^{2}).
\end{equation}
To extrapolate the energy gaps, we take the finite size correction to $V^{n}_{m}$ for each system size.\cite{Morf2002}

We first report the results obtained by the exact diagonalization method at $\nu =7/3$.\cite{d'Ambrumenil1988}
Although the maximum value of $\alpha$ given by the spin orbit interaction is 0.5,
we increase $\alpha$ for $\alpha \geq   0.5$ to see the relation between the Laughlin state and the mixed state written by eq. (5).
Figure 1 shows the energy gaps for $0.2 \leqq \alpha \leqq  1.0 $.$ ^{27)}$ 
At $\alpha = 1.0$, the value of the extrapolated gap obtained by the best linear fit is 0.101.
This value is equal to the result of the previous work.\cite{Note1}
The energy gap first increases and then decreases with the increase of $\alpha$.
The $\alpha$ which gives the maximum energy gap is around 0.55.
We discuss this behavior by considering the relation between the pseudopotential and the energy gap in the Laughlin state.
The energy gap of the Laughlin state is roughly proportional to the difference of $V_{1}^{1}$ and $V_{3}^{1}$ obtained by eq. (11).
$V_{1}^{1}-V_{3}^{1}$ is written as a function of $\alpha$ by \\
\begin{equation}
V_{1}^{1}-V_{3}^{1} = -0.25963\cos ^{4}(\frac{\pi}{2}\alpha)+0.19386\cos^{2}(\frac{\pi}{2}\alpha)+0.16616 .
\end{equation}
 We compare $V_{1}^{1}-V_{3}^{1}$ with the value of the extrapolated energy gaps in Fig.2.
 This figure shows that the $\alpha$ dependence of the energy gap and the $\alpha$ dependence of $V_{1}^{1}-V_{3}^{1}$ are similar.
 This result also means that the GS for $0.2 \leqq \alpha\leqq  1.0$ is expected to be the Laughlin state.
 Therefor the effect of the spin-orbit interaction on the Laughlin state is featured
 by the change in the pseudo potential.
 We therefor conclude that the Laughlin state is stabilized by the spin-orbit interaction.\\         
\begin{figure}
\begin{center}
\scalebox{0.5}{\includegraphics[width=180mm,angle=0]{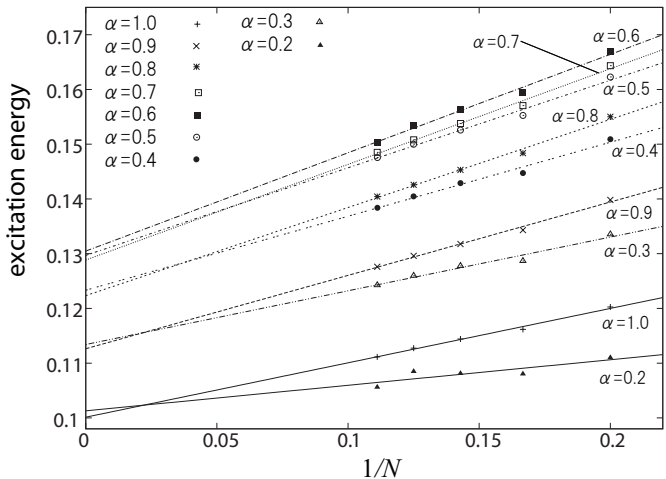}}
\end{center}
\caption{Energy gaps at $\nu =7/3$. The straight lines denote the best linear (in $1/N$) fit to the data points. N is the number of electrons.}
\end{figure}
 \begin{figure}[t]
\begin{center}
\scalebox{0.5}{\includegraphics[width=180mm, angle=0]{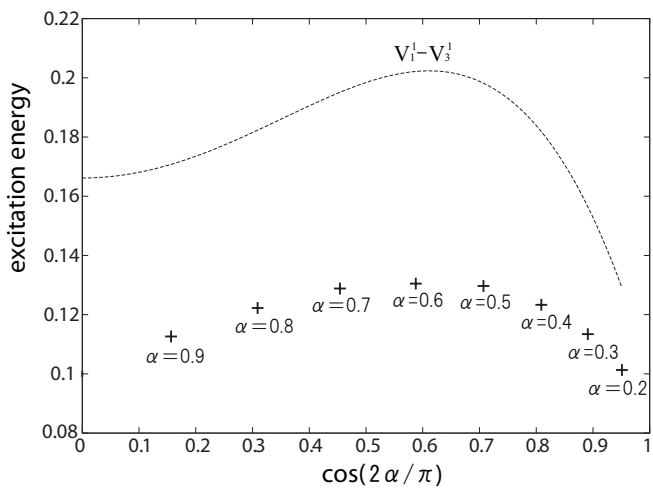}}
\end{center}
\caption{ The cross marks represent extrapolated energygaps. The broken
curve denotes $V_{1}^{1}-V_{3}^{1}$. The horizontal axis is  $\cos(\frac{\pi}{2}\alpha)$.}
\end{figure}

Next we investigate the GS at $\nu =5/2$ to see the effect of the spin-orbit interaction on the Pf state.
In Fig.3, we show our results of the energy gaps for $0 \leqq \alpha \leqq  0.3 $. Half the sum of quasiparticle
and quasihole excitation energies are plotted as a function of $1/N$, where $N$ is number of electrons in the system.
At $\alpha = 0$, the plotted data are well fitted by a straight line in $1/N$. 
Our result at $\alpha = 0$ calculated by the DMRG method is 0.025.
This value is equal to the previous result calculated by exact diagonalization method.\cite{Morf2002}
The results in Fig.3 show that the energy gap first increases and then decreases with the increase of $\alpha$ from $\alpha = 0$.

We discuss the $\alpha$ dependence of the energy gap by considering the phase diagram of the GS.
 In general, the value of the energy gap depends on the distance from the phase boundary.
 In the previous studies,\cite{Morf1998,Wang2009,Storni2010} an extra increase of the pseudopotential induce the transition from the Pfaffian state to the CFL phase.
 The increase of the pseudopotential corresponds to the increase of $\alpha$ for $0 \leqq \alpha \leqq  0.3 $.  
 The GS at $\alpha = 0.1$ is away from the stripe phase and the CFL phase,  
 consequently the energy gap at $\alpha = 0.1$ is larger than the gap at $\alpha = 0$.\\

\begin{figure}[thbp]
\begin{center}
\scalebox{0.5}{\includegraphics[width=180mm, angle=0]{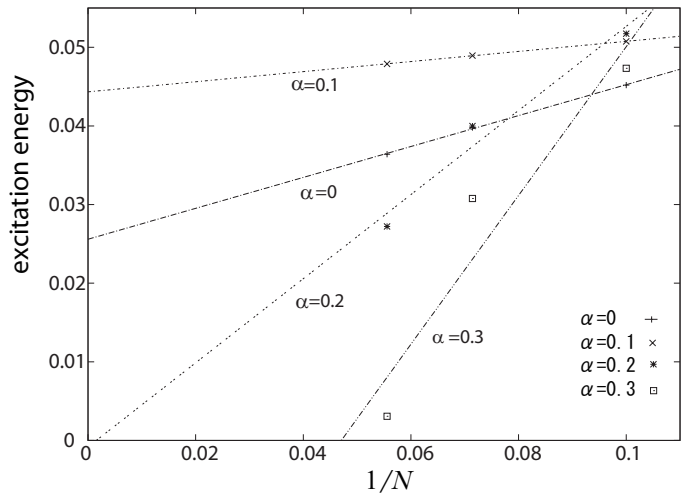}}
\caption{Energy gaps at $\nu =5/2$. The straight lines denote the best linear (in $1/N$) fit to the data points. N is the number of electrons. }
\end{center}
\end{figure}
\:In order to further study of the GS around $\alpha = 0.1$, we next investigate the pair-correlation functions.
The pair-correlation function is defined by
\begin{equation}
g(r)=\frac{1}{2\pi\sin(\theta)}\sum _{i<j}\langle\delta(\theta+\theta_{j}-\theta_{i})\rangle .
\end{equation}  
Figure 4 shows the results of the pair-correlation functions for $N$ =18.
 The curves for $\alpha = 0$ , 0.1 and 0.2 are almost indistinguishable.
 However, the correlation function at $\alpha =0.3$ looks that of the CFL state.
 Precisely in a short distance ($r<2$), the correlation functions for $\alpha \leq  0.2$ have shoulder structure
 which is a signature of the Pf state but
 the correlation function at $\alpha =0.3$ does not have such a Pf-like structure.
 In a long distance ($r>4$), the amplitude of oscillation of the correlation function at $\alpha =0.3$ is smaller than other correlation functions.
 From these differences, we expect that the increase in $\alpha$ would induce the phase transition.
 This result agree with our expectation.\\  
\begin{figure}[hbp]
\begin{center}
\scalebox{0.5}{\includegraphics[width=180mm, angle=0]{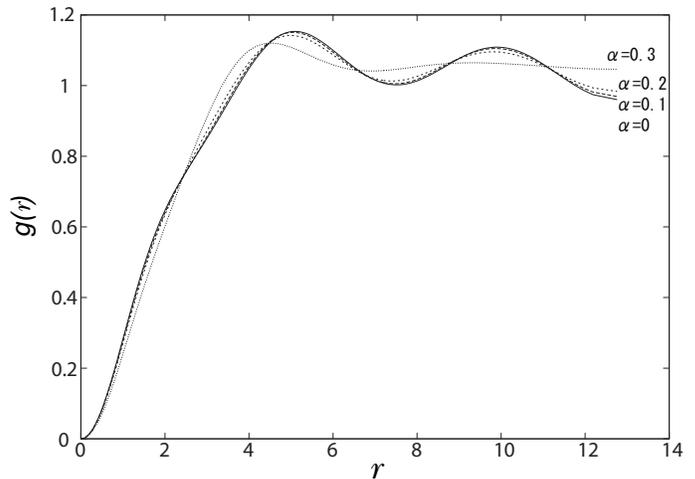}}
\end{center}
\caption{The pair correlation functions at $\alpha  = 0$, 0.1, 0.2 and 0.3 for $N=18$.}
\end{figure}

To confirm the Pf-like state around $\alpha = 0.1$,
we finally calculate the overlap between the exact ground state and the Pf wave function for $N =6$. 
The Pfaffian wave function on the spherical geometry is written by

\begin{equation}
\Psi_{\rm Pf}={\rm Pf}\Big(\frac{1}{u_{i}v_{j}-v_{i}u_{j}}\Big)\prod _{i<j}(u_{i}v_{j}-v_{i}u_{j})^2,
\end{equation}
here,
$u=\cos(\frac{\theta}{2})e^{i\phi}$, and $v=\sin(\frac{\theta}{2})e^{-i\phi}$, and we choose a gauge such that $A=(\frac{N_{\Phi}}{qR})\cot(\theta)\textrm{\boldmath $\phi$}   $
where $\theta,\phi$ are the polar coordinates on the unit sphere with $N_{\Phi}$ units of flux passing through the sphere.
 The value of the overlap is defined by $|\langle \Psi_{\rm exact}|\Psi_{\rm Pf}\rangle |$.
 The overlap at $\alpha=0$ is 0.866 and it increases with the increase in $\alpha$.
 The maximum value of the overlap is 0.901 at $\alpha = 0.15$.
 This behavior is consistent with that of the energy gap.
 We therefor confirmed that the GS at $\alpha = 0.1$ is characterized by the Pf state which is stabilized by the spin-orbit interaction.\\

In summary,
we have investigated the effect of the spin-orbit interaction on the Laughlin state and the Pf state.
We have introduced the parameter $\alpha$ which is 0 in the absence of the spin-orbit interaction and
it increases with the increase of the spin-orbit interaction.  
For the Laughlin state at $\nu = 7/3$, we find that the energy gap is increased by the spin-orbit interaction.
The enhancement of the energy gap is reasonably understood by the Haldene's pseudopotential written in eq. (11).
For the Pf state at $\nu = 5/2$,
we have calculated the energy gap, the correlation functions, and the overlap as a function of $\alpha$.
Both the extrapolated energy gap and the overlap first increase and then decrease with the increase of $\alpha$ from $\alpha=0$.
The energy gap has the maximum value at around $\alpha = 0.1$ and the maximum of the overlap is 0.901 at $\alpha = 0.15$ for $N=6$.
The pair-correlation functions for $0 \leqq  \alpha \leqq 0.2$ are almost the same.
These correlation functions are characterized by the Pf-like state,
and we can expect that the GS is the Pf state.  
The pair-correlation functions at $\alpha = 0.3$ is CFL-like state.
This means that the increase in $\alpha$ induce the phase transition. 
From above results, we conclude that the Pf state is stabilized by the increase of the spin-orbit interaction.
Spin-orbit coupling linery depends on the interface electron field which is controlled by the gate voltage.\cite{Nitta1997}
This means that the FQH state is stabilized by the gate voltage which is applied perpendicular to the quantum well.\\

\textbf{Acknowledgements}\\
\:TI thanks Dr. T. Higashi for helpful discussions and advice.
The present work is supported by Grant-in-Aid No. 18684012 and  No. 20740167 from MEXT, Japan.

\noindent

\end{document}